# Laser Cavity-Soliton Micro-Combs

Hualong Bao[1], Andrew Cooper[1], Maxwell Rowley[1], Luigi Di Lauro[1], Juan Sebastian Totero Gongora[1], Sai T. Chu[2], Brent E. Little[3], Gian-Luca Oppo[4], Roberto Morandotti[5,6,7], David J. Moss[8], Benjamin Wetzel[1], Marco Peccianti[1] and Alessia Pasquazi[1*]

*[1]Emergent Photonics (Epic) Lab, Dept. of Physics and Astronomy, University of Sussex, BN1 9QH, UK*
*[2]City University of Hong Kong, Tat Chee Avenue, Hong Kong, China*
*[3]Xi'an Institute of Optics and Precision Mechanics, Chinese Academy of Science, Xi'an, China*
*[4]SUPA, Department of Physics, University of Strathclyde, Glasgow, United Kingdom*
*[5]INRS-EMT, 1650 Boulevard Lionel-Boulet, Varennes, Québec, Canada J3X 1S2*
*[6]Institute of Fundamental and Frontier Sciences, University of Electronic Science and Technology of China, Chengdu 610054, Sichuan, China*
*[7]ITMO University, St. Petersburg 199034, Russia*
*[8]Centre for Microphotonics, Swinburne University of Technology, Hawthorn, VIC 3122, Australia*

**Corresponding author: a.pasquazi@sussex.ac.uk*

**The field of micro-cavity based frequency combs, or 'micro-combs'[1,2], has recently witnessed many fundamental breakthroughs[3-19] enabled by the discovery of temporal cavity-solitons, self-localised waves sustained by a background of radiation usually containing 95% of the total power[20]. Simple methods for their efficient generation and control are currently researched to finally establish micro-combs as out-of-the-lab widespread tools[21].**

**Here we demonstrate *micro-comb laser* cavity-solitons, an intrinsically highly-efficient, background free class of solitary waves. Laser cavity-solitons have underpinned key breakthroughs in semiconductor lasers[22,23] and photonic memories[24-26]. By merging their properties with the physics of both micro-resonators[1,2] and multi-mode systems[27], we provide a new paradigm for the generation and control of self-localised pulses in micro-cavities. We demonstrate 50 nm wide soliton combs induced with average powers *one order of magnitude* lower than those typically required by state-of-the-art approaches[26]. Furthermore, we can tune the repetition-rate to well over a megahertz with no-active feedback.**

Optical frequency combs based on micro-cavity resonators, also called 'micro-combs', offer the promise of achieving the full capability of their bulk counterparts in an integrated footprint[1, 2]. They have enabled major breakthroughs in spectroscopy[3,4], communications[5,6] microwave photonics[7], frequency synthesis[8], optical ranging[9,10], quantum sources[11, 12] and metrology[13,14]. Of particular importance was the discovery of temporal cavity-solitons in micro-cavities[15-19].

Temporal cavity-solitons[2,15-20,26] are an important example of dissipative solitons – self-confined waves in space or time that balance the nonlinear phase-shift with wave dispersion in systems featured by gain and loss[28]. A practical use of these pulses for micro-combs, however, still requires overcoming important challenges. As they exist in bistable systems as localised states upon a continuous-wave (CW) background[2,15-20,26], they show very limited conversion efficiency, usually as low as a few per cent[20]. Furthermore, controlling their fundamental parameters, such as their repetition-rate, has also posed a challenge. Specifically, tuning their repetition-rate currently requires either complex methods involving fast detection, microwave signal processing and fast cavity actuation, or novel approaches, such as pulsed[29] or counter-propagating[30] pumping, as well as heterodyning with coupled micro-resonators[31]. Finding a solution to these fundamental problems is attracting important technological efforts[21].

Here, we approach these challenges demonstrating a distinct class of solitary pulses – temporal ***laser*** cavity-solitons – in micro-combs. Laser cavity-solitons[22-26] have been largely studied in spatial configurations such as semiconductor lasers[22], where they enabled breakthroughs such as all-optical reconfigurable memories[23]. More recently, they have been observed in both temporal[25] and spatio-temporal[32] contexts. Different to externally-driven cavity-solitons used in micro-combs, which are sustained by the energy of the pumping background, laser cavity-solitons receive energy directly from the gain of a lasing medium. For this reason, they exist without any background light and are intrinsically the most energy-efficient class of cavity-solitons.

By nesting a Kerr micro-resonator in a fibre loop with gain, we harvest the intrinsic capability of laser cavity-solitons, demonstrating that they can be used to achieve highly-efficient, broadband micro-comb generation. We obtained background-free pulses, with bandwidth over 50 nm, excited with average powers less than ***6% of the threshold*** for CW-pumped micro-combs in an equivalent resonator. Furthermore, by exploring the properties of multi-mode systems, recently investigated for spatio-temporal mode-locking[27] and spatial beam self-cleaning[33], we show that the repetition-rate of our pulses can be adjusted by reconfiguring simple parameters such as the laser cavity-length. With no-active control, we succeed in modifying their repetition-rate by more than a megahertz.

Our architecture is inspired by previous experiments[2,34-35] and, in particular, by the concept of filter-driven four-wave mixing laser[34]. While those results achieved stable pulse generation, the pulse widths were limited to the picosecond regime[2]. Furthermore, the role of the frequency-detuning and free-spectral range mismatches between the two cavities was neither understood nor exploited. Here we show that these parameters are indeed critical to realise this class of solitary pulses.

The principle of operation is shown in Fig. 1a. A nonlinear micro-cavity (cavity 'a') is embedded within a longer fibre-amplifying cavity (cavity 'b'). These two simple elements are sufficient for generating laser cavity-solitons[22,36]: the pulse propagating in the fibre loop, spectrally limited by the laser-gain bandwidth, sustains the existence of a spectrally-broadened pulse within the Kerr micro-cavity. These pulses do not require fast saturable absorption, in sharp contrast to passively mode-locked lasers that intrinsically need a gain-dampening mechanism. This is a typical aspect for cavity-solitons, due to their nature of localised pulses in bistable systems[26].

Temporal cavity-solitons can be effectively modelled using mean-field approaches, such as the Lugiato-Lefever equation[15], which is widely employed for micro-combs, where the field in the micro-cavity is described as a pulse propagating in time, along with a spatial coordinate periodically looped within the micro-resonator length. Here we build a set of multi-component (or 'vectorial') laser mean-field equations[27,33] by coupling together the micro-cavity field with the main-cavity 'super-mode' fields. (Supplementary/Methods). A super-mode is an optical radiation formed by a set of equally-spaced modes of the main-cavity, whose relative spacing (in frequency) is set by the micro-cavity free-spectral range (FSR) $F_a$. Quantitively, the m$^{th}$ resonance $f_m^{(b)}$of a super-mode can be linked to the micro-cavity resonances $f_m^{(a)}$ by the relation

$$f_m^{(b)} = f_m^{(a)} + F_b(q - \Delta + m\,\delta), \tag{1}$$

where $F_b$ is the main-cavity FSR and $q$ is an integer defining the order of the super-mode (see Fig. 1b). In general, the $q$-order super-mode is frequency detuned with respect to the micro-cavity resonance by $(\Delta - q)\,F_b$, where $\Delta$ is the cavity-frequency offset, normalised against $F_b$. The key features of the laser are determined by the leading-order super-mode, defined for $q = 0$, which possesses the largest spectral overlap with the micro-cavity resonances. Higher-order super-modes ($q \neq 0$) typically experience greater coupling losses. Because $F_b$ is not necessarily an integer divisor of $F_a$, we introduce the variable $\delta$, normalised against $F_b$, representing the FSR detuning.

Two numerical examples of linear and solitary propagation are reported in Fig. 2. We use a spatial coordinate periodically closed over the micro-cavity length because the temporal waveform of every super-mode is periodic with the micro-cavity round-trip time $T_a$, with a period slightly detuned by $\delta$ (Eq. (1)). The parameters $\Delta - q$ and $\delta$ play the role of the frequency and group velocity mismatches between the micro-cavity and super-mode fields. A key property of solitary waves in vectorial equations, also recently shown for spatio-temporal mode-locking[27], is that all the coupled fields *lock* to a single group velocity (or repetition-rate detuning) $v$ (Fig. 2a and b) and carrier frequency offset $\phi$ (Fig. 2c). Practically, the system provides a well-defined soliton comb, with the $n^{th}$ frequency tooth $f_n^{(S)}$ expressed as

$$f_n^{(S)} = f_n^{(a)} + F_b(\,\phi - n\,v). \tag{2}$$

The parameters $\phi$ and $v$ are selected by $\Delta$ and $\delta$, together with the normalised saturated gain $g$ (Methods/Supplementary).

The measurements of two soliton micro-combs at different intra-cavity powers are reported in Figure 3a-d. The spectra exhibit a bandwidth (up to 50 nm) comparable to the cavity-solitons observed in resonators with similar dispersion properties[9,18,19] and, together with the related autocorrelations (Fig. 3a, c and inset), are in excellent agreement with theory. Like those pulses, our solitary waves are achieved in red-detuned cavities, as revealed by the relative position of the oscillating modes within the micro-cavity resonance (Fig. 3b and d). These plots, obtained by intra-cavity laser-scanning spectroscopy (Methods), also reveal a small, blue-detuned frequency found in the central resonances but absent in the comb wings. We attribute it to a small CW perturbation in the first-order super-mode (see Supplementary). The presence of perturbations with a frequency distinct from the stationary solution is in stark contrast to externally-driven micro-cavity systems, where all the co-existing solutions[26] (cavity-solitons and their perturbations) are intrinsically locked to the driving-field frequency.

These experiments demonstrate the inherently higher efficiency of our class of cavity-solitons compared to Lugiato-Lefever ones. In both cases, the cavity-solitons require bistability, i.e. the contemporary presence of both a high- and a low-energy stationary state upon which the soliton is formed. For Lugiato-Lefever solitons, the low-energy state is a CW background that, in the comb spectrum, results in a dominant, high-power mode which contains most of the comb energy (95% and 50%, respectively, for bright and dark solitons[20]). Because our low-energy state is zero (see Fig. 4a and Methods/Supplementary), as typical in systems with gain, the most energetic comb mode in our experiments contains less than 25% of the total energy, with a theoretical limit of 4%. Moreover, Lugiato-Lefever solitons feature a minimum power excitation threshold above which the Kerr nonlinearity induces the bistability, i.e. two CW states necessary to their existence. Our laser cavity-solitons, which require a zero background and a single CW state, do not have a minimum excitation power (Fig. 4b). By comparing experiments and theory, we find that the peak powers injected into the micro-cavity for the experimental cases in Fig. 3a and b are, respectively, 40% and 50% of the input power *threshold* of a Lugiato-Lefever soliton in the same micro-resonator. This is a remarkable result in terms of generation efficiency, not only because Lugiato-Lefever solitons are usually excited well above their threshold, but also because our injected field is pulsed. Hence, the injected average powers to the micro-resonator in our two experiments are, respectively, as low as 5% and 6% of the Lugiato-Lefever soliton power threshold (Supplementary).

To demonstrate the capability of changing the soliton repetition-rate with simple methods, we varied the main-cavity FSR detuning $\delta$ with a delay line that modified the fibre cavity length and, hence, the mode-spacing $F_b$. Gain and loss were also adjusted to maintain the solitary state. Fig. 5 shows the repetition-rate variations for three stable soliton combs, each with different power (Fig. 5a). We measured by laser-spectroscopy the frequency position of the comb modes against the mode number. We then calculated the best-fit for the first case (Comb 1) and subtracted to the frequency positions for the three cases, obtaining the residual frequency versus mode number in Fig. 5b. This shows a change in repetition-rate of over a MHz. The theoretical results (Supplementary) demonstrate that, by changing $\delta$ in the experimental range, the soliton is maintained stable and modifies its velocity.

In conclusion, we report the observation of temporal laser cavity-solitons in optical micro-combs. Our results merge the powerful physics of optical Kerr micro-combs and their ability to generate large bandwidths (potentially allowing the use of dispersive wave coupling[13] to expand the spectrum further) with the unique properties of laser cavity-solitons and multi-mode systems. In contrast to conventional coherently-driven cavity-solitons[15-21,29-31], this new class of cavity-solitons is intrinsically background free and characterised by the absence of a minimum power excitation threshold, making them extremely energy efficient. Furthermore, thanks to a tailored two-cavity configuration, crucial properties, such as the repetition-rate, can be controlled with simple elements such as a delay line. Finally, we achieved these localised states by manual adjustment of the fibre cavity parameters, such as cavity length, gain current and polarisation losses, in a similar fashion to passively mode-locked lasers: this approach enables the use of powerful methods, such as genetic algorithms, that have been instrumental in achieving adaptive control of the soliton properties and self-starting operation in passive mode-locking[37].

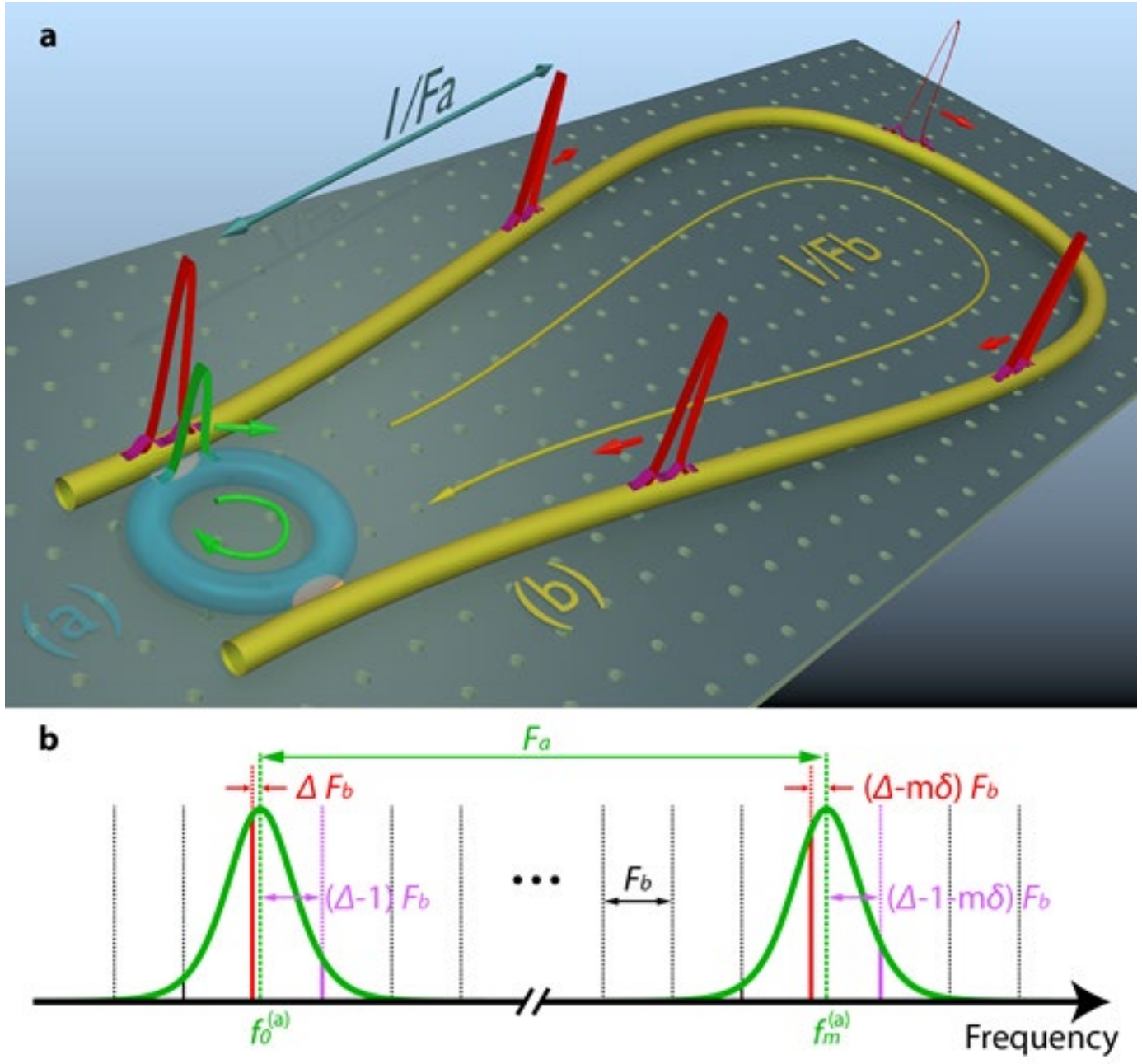

**Figure 1. Principle of operation of micro-comb laser cavity-soliton formation**. **a.** A short pulse (green) propagates in the micro-cavity (blue) sustained by a longer pulse (red) and a weak higher-order ‘super-mode pulse’ (purple) in the amplifying loop (yellow). **b.** Cold-cavity spectral distribution. Micro-cavity resonances are depicted in green, amplifying-cavity resonances are in black, with leading and first-order super-modes highlighted in red and purple, respectively. The normalised frequency offset between the central frequency of the leading super-mode and the micro-cavity resonance is $\Delta$; similarly, the frequency offset is $\Delta$-1 for the central frequency of first-order super-mode. The variable $\delta$ is the normalised FSR detuning, appearing when the two cavities are not commensurate.

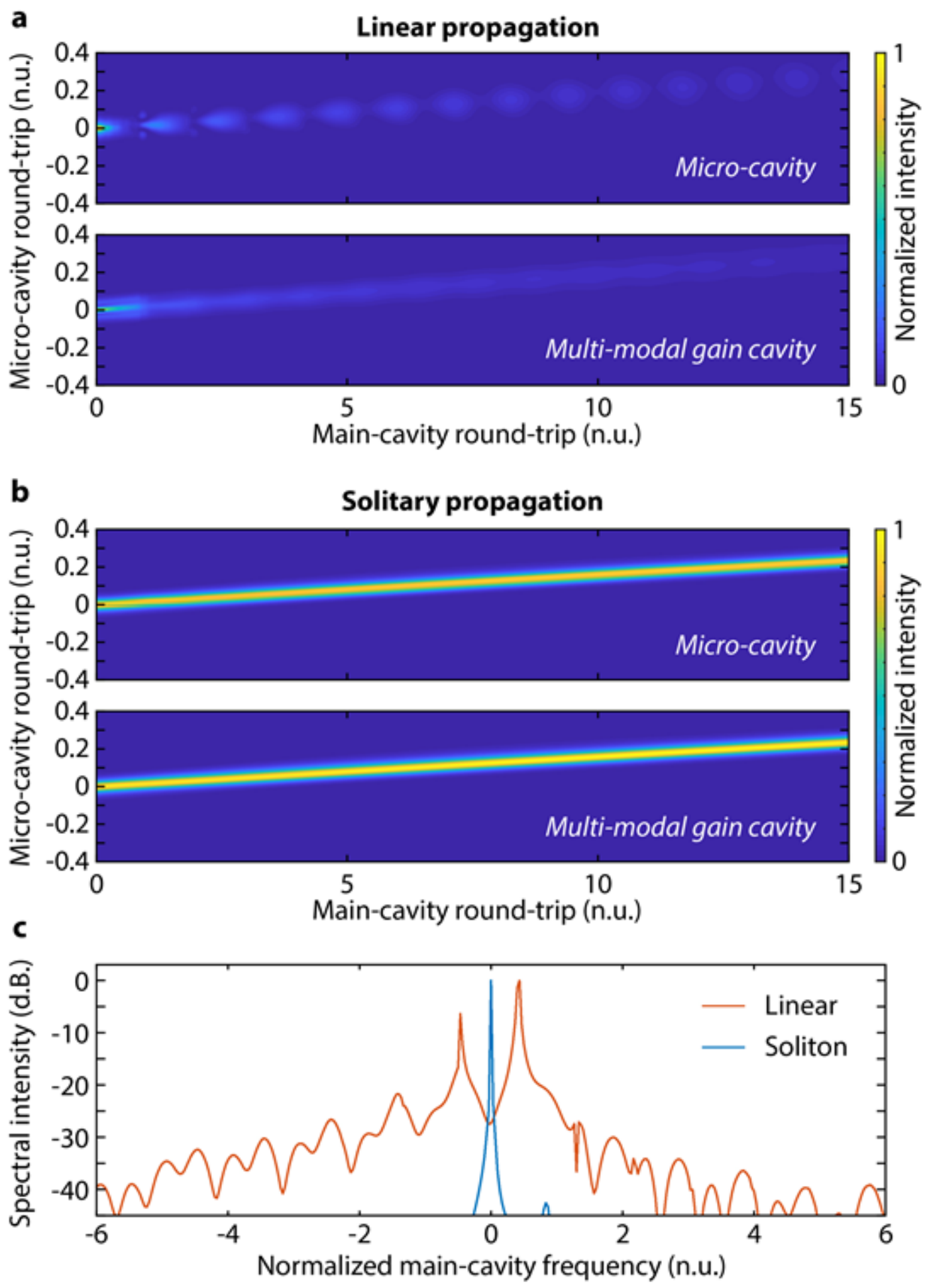

**Figure 2. Theoretical propagation of linear and solitary pulses.** Micro-cavity and gain cavity have a group velocity mismatch $\delta$ = *0.03.* **a.** Evolution of the micro-cavity and multi-modal amplifying-cavity fields in the linear case. The group velocity mismatch causes a periodical decoupling of the fields. **b.** Solitary propagation: both fields lock to the same group velocity $v$. **c.** Equivalent spectral distribution of the super-modes within a resonance of the micro-cavity (Supplementary). Such a spectrum, for the linear case (orange), highlights the presence of many frequency components, one for every super-mode with frequency offset $\Delta - q$. In the case of solitary propagation (blue), conversely, all the modes *lock* to the frequency $\phi$.

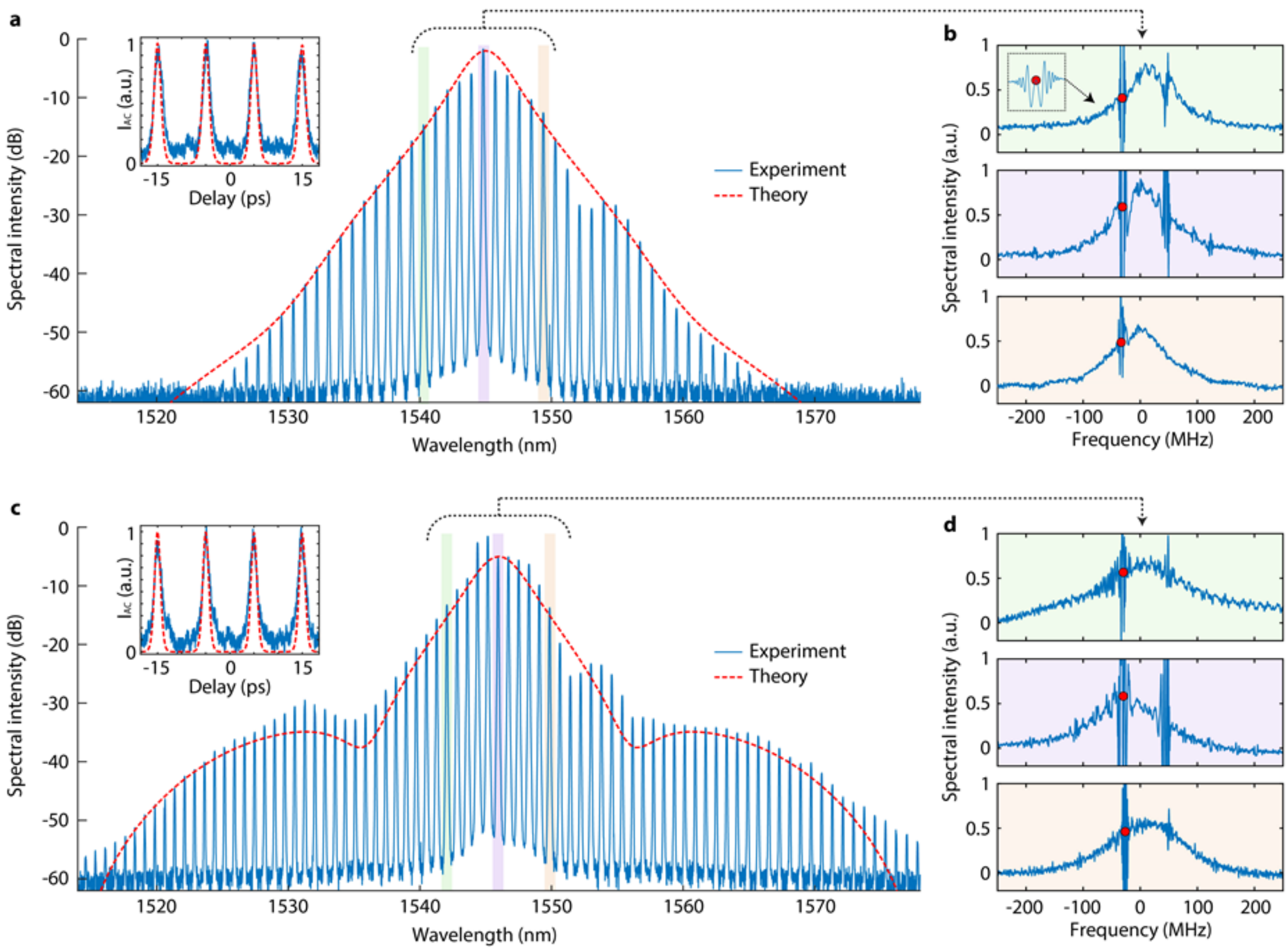

**Figure 3. Temporal laser cavity-soliton measurement. a.** Soliton generation with 20 mW intra-cavity power: spectrum (in logarithmic scale) and autocorrelation (inset). The experimental measurements are plotted in blue and directly compared to the theoretical solitary state (red, fit parameters: Δ = 0.49; g= 0.1). **b.** Intra-cavity spectrum (blue), evidencing the lasing modes (red dots, appearing for lower frequencies than the central resonance frequency) within each micro-cavity resonance. The three plots correspond to different comb wavelengths highlighted in panel (a) by the colour shading. **c, d.** The same measurements are at higher fibre gain, leading to a 30 mW intra-cavity power and larger comb spectrum. Fit parameters are Δ =0.47; g= 0.14.

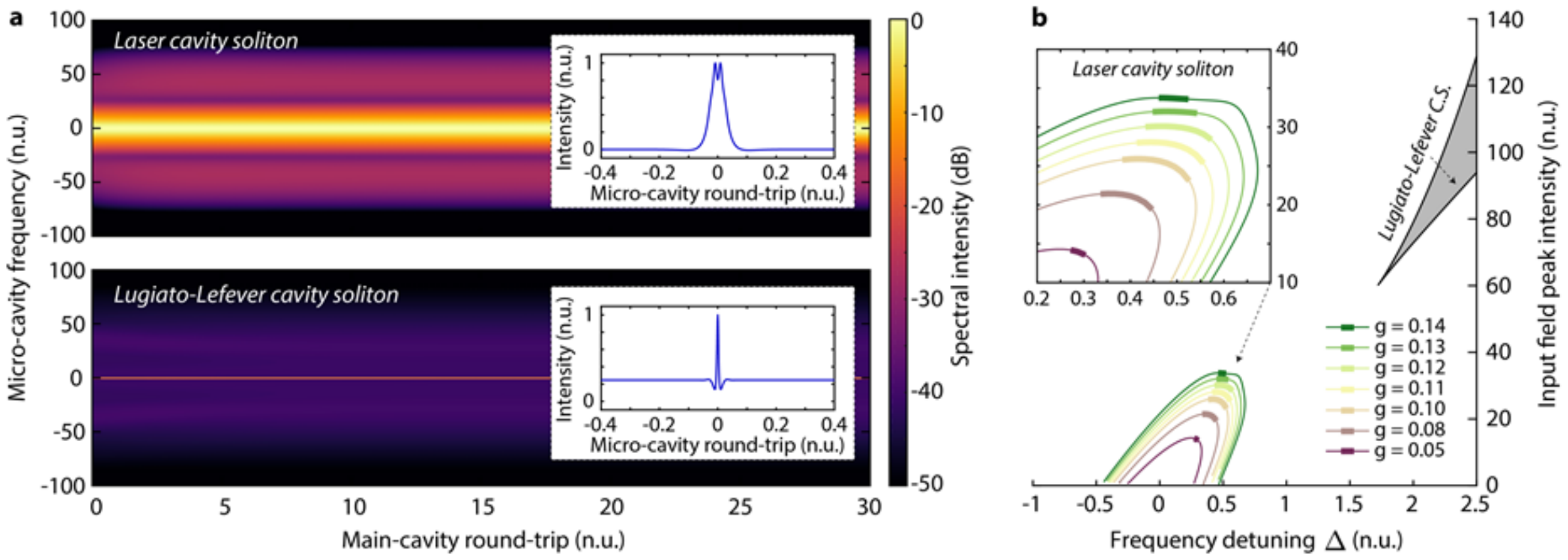

**Figure 4. Temporal laser cavity-soliton and Lugiato-Lefever cavity-soliton comparison. a.** Spectral evolution of a laser cavity-soliton (Δ =0.47; g= 0.14.) and a Lugiato-Lefever cavity-soliton at the power threshold, for the same resonator property, and pulses profiles (inset) showing that laser cavity-soliton comb lines generally possess higher power spectral density. **b.** Plot of laser cavity-soliton input field peak power versus normalised offset Δ, calculated for various gain values (g=0.05 to 0.14 for plots from purple to green). Thick lines in the zoomed inset mark the stable self-localised solutions. Grey region marks the region of existence of the Lugiato-Lefever bright solitons.

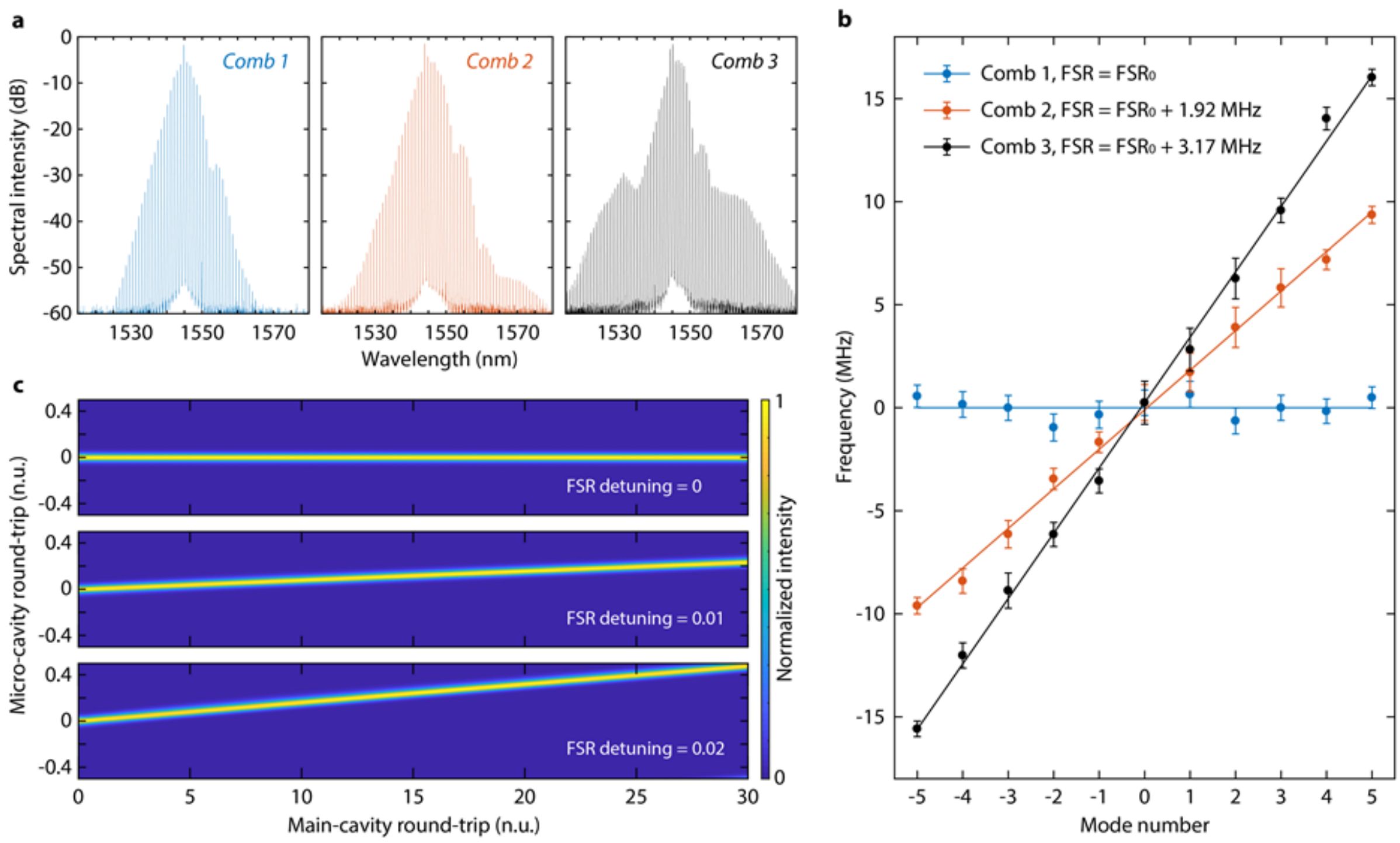

**Figure 5. Temporal laser cavity-soliton repetition-rate control. a.** Spectra. The fibre cavity length is changed within a broad range of 150 μm. Gain and losses had been readjusted to maintain the soliton state; intra-cavity powers are 20 mW, 25 mW and 30 mW, in blue, orange and black, respectively. **b.** Residual frequency shift against mode number with respect to the best-fit for the Comb 1 case (blue). Combs 2 and 3 show a change in FSR and, hence, repetition-rates of 1.9MHz and 3.2MHz with respect to Comb 1. **c.** Calculated propagation of a stationary solitary solution when $\Delta$ =0.49 and g= 0.1, for $\delta = 0$; 0,01; 0.02. The solitary wave is maintained in all three cases.

**Methods**

**Experimental Setup**

The experimental setup is built around a silicon oxynitride integrated resonator with FSR~48.97 GHz and linewidth $\Delta F_A$ <140MHz (within the experimental measurement uncertainties), corresponding to a Q factor of ~1.3 million. The micro-ring is inserted in an anomalous dispersion, polarisation-maintaining Ytterbium-Erbium co-doped fibre-cavity. This cavity comprises a delay line, a 10 nm tuneable bandpass filter (resulting, when matched to the amplifier spectral response, in a Gaussian linewidth of $\Delta F_F$ =650 MHz), polarisation optics and an optical isolator. The FSR of the cavity is about 77 MHz. The use of a full polarisation-maintaining loop prevents any effect related to fast gain saturation. Note that the fibre amplifier has a recovery time in the order of 10 ms[38], well above the soliton repetition-rate and main-cavity round-trip. The intra-cavity lines are measured by laser-scanning spectroscopy (Supplementary) . The scanning laser is calibrated with a 6.95 MHz Mach-Zender and by beating it with a 250MHz reference comb[39]. The micro-comb output is extracted at the drop-port of the micro-ring and is characterised with a second harmonic generation, background-free, non-collinear autocorrelator, optical spectrum-analyser and radio-frequency detection, obtained with a fast oscilloscope. The soliton intra-cavity energy is measured at the drop-port of the micro-resonator. The soliton cases presented in the experiments show the presence of two pulses in the cavity, as the lines of the soliton spectra are spaced every other mode. Our solitons are obtained with a modest 150 mW intra-cavity laser power measured at the output of the fibre amplifier before all the system losses.

**Model**

The main features of the laser can be obtained by a simple mean-field model (see Supplementary for the complete derivation) that, in its normalised form, reads:

$$\partial_t a + \frac{i\zeta_a}{2}\partial_{xx}a + i\,|a|^2 a = -\kappa a + \sqrt{\kappa}\sum_{q=-N}^{N} b_q \quad (3)$$

$$\partial_t b_q + \delta\partial_\tau b_q + \frac{i\zeta_b}{2}\partial_{xx}b_q - 2\pi\, i\,(\Delta - q)b_q = -\sigma\partial_{xx}b_q + g\, b_q - \sum_{p=-N}^{N} b_p + \sqrt{\kappa}a, \quad (4)$$

where $a$ and $b_q$ are the optical field envelopes for the microresonator and amplifying cavities, respectively, and are expressed as a function of the normalised propagation time $t$ and space coordinate $x$. Here we have considered the generic interaction with 2N+1 super-modes $b_q$, for $|q| \leq N$; the mode with $q=0$ corresponds to the leading mode.

The time $t$ accounts for the propagation over different round-trips and is normalised against the main-cavity round-trip $T_b=1/F_b$=12.5 ns. The space coordinate $x$, defined for $|x|<1/2$, is associated to the frame moving with the pulse and is normalised against the micro-cavity round-trip length, which corresponds to a round-trip time $T_a=1/F_a$ =20 ps.

The left- and right-hand sides of the equations contain the conservative and dissipative terms: $\zeta_{(a,b)} >0$, $\Delta$ and $\delta$ are the normalised coefficients for the cavity (anomalous) dispersions, the cavity-frequency offset and the group velocity mismatch. The latter term considers the effective FSR detuning between the two cavities as in Eq. (1); $\kappa$, $g$ and $\sigma$ represent the coupling, saturated gain and bandwidth of the spectral-filtering, respectively.

Specifically (see Supplementary for the derivation), the normalised coupling parameter is $\kappa = \pi \Delta F_A T_b \approx 2\pi$, being the micro-cavity linewidth $\Delta F_A$ approximately twice the FSR of the main-cavity (in the experimental condition). The normalised dispersions are $\zeta_{(a,b)} = -\beta_{(a,b)} v_{(a,b)} T_b / T_a^2$ for which $v_{(a,b)}$ and $\beta_{(a,b)}$ are the group-velocities and group velocity dispersions of the two cavities, respectively. In the simulations, we used $\zeta_a = 1.25 \times 10^{-4}$, $\zeta_b = 3.5 \times 10^{-4}$ , obtained with values $|\beta_a|$≈-20 ps$^2$/km and $|\beta_b|$≈-60 ps$^2$/km (within our experimental constrains). We used a gain bandwidth $\sigma = (2\pi T_a \Delta F_F)^{-2} \approx 1.5 \times 10^{-4}$, based on a 650 THz intra-cavity spectral filter. The gain $g$, considered as the saturated-gain of the amplifier, is normalised against the main-cavity length and, together with $\Delta$ and $\delta$ in Eq. (1), is an adjustable parameter in our numerical datasets.

The stationary states are defined as $a(t,x) = a_S(x - vt)\exp[2\pi i\phi t]$, $b_q(t,x) = b_{q,S}(x - vt)\exp[2\pi i\phi t]$, where the normalised frequency offset $\phi$ and the normalised velocity $v$ are as in Eq. (2). Solitary solutions are found by numerical continuation considering 11 super-modes (i.e. with N=5), while stability is investigated with linear perturbation analysis and propagation considering 61 super-modes (i.e. with N=30).

**Acknowledgment**

We acknowledge the support of the U.K. Quantum Technology Hub for Sensors and Metrology, EPSRC, under Grant EP/M013294/1, of the Marie Curie Action MC-CIG and IIF REA grant 630833 and 327627. M.P acknowledge additional support by EU-H2020, research and innovation programme (725046). B.W. acknowledges the support from the People Programme (Marie Curie Actions) of the European Union's FP7 Programme under REA grant agreement INCIPIT (PIOF-GA-2013-625466).

R.M. acknowledges funding by the Natural Sciences and Engineering Research Council of Canada (NSERC) through the Strategic, Discovery, and Acceleration Grants Schemes, by the MESI PSR-SIIRI Initiative in Quebec, by the Canada Research Chair Program, as well as additional support by the Government of the Russian Federation through the ITMO Fellowship and Professorship Program (grant 074-U 01) and by the 1000 Talents Sichuan Program.

# Laser Cavity-Soliton Micro-Combs

## Supplementary Materials

Hualong Bao[1], Andrew Cooper[1], Maxwell Rowley[1], Luigi Di Lauro[1], Juan Sebastian Totero Gongora[1], Sai T. Chu[2], Brent E. Little[3], Gian-Luca Oppo[4], Roberto Morandotti[5,6,7], David J. Moss[8], Benjamin Wetzel[1], Marco Peccianti[1] and Alessia Pasquazi[1*]

*[1]Emergent Photonics (Epic) Lab, Dept. of Physics and Astronomy, University of Sussex, BN1 9QH, UK*
*[2]City University of Hong Kong, Tat Chee Avenue, Hong Kong, China*
*[3]Xi'an Institute of Optics and Precision Mechanics, Chinese Academy of Science, Xi'an, China*
*[4]SUPA, Department of Physics, University of Strathclyde, Glasgow, United Kingdom*
*[5]INRS-EMT, 1650 Boulevard Lionel-Boulet, Varennes, Québec, Canada J3X 1S2*
*[6]Institute of Fundamental and Frontier Sciences, University of Electronic Science and Technology of China, Chengdu 610054, Sichuan, China*
*[7]ITMO University, St. Petersburg 199034, Russia*
*[8]Centre for Microphotonics, Swinburne University of Technology, Hawthorn, VIC 3122, Australia*

**Corresponding author: a.pasquazi@sussex.ac.uk*

## Derivation of the Model Equations

Our experimental configuration can be described by a system comprising two travelling-wave resonators, as depicted in Fig. S1a. Such a system features a Kerr cavity ('a', red circle) nested in an amplifying main-cavity ('b', black loop). In our experiments, such cavities are, respectively, a micro-ring resonator and a fibre-loop with gain. The amplifying-cavity is effectively an open loop, linked to the input and output ports of the ring. For simplicity, we model it as a closed cavity, adding an effective length of approximately half the circumference of the ring between these two connection points. The main-cavity field at the input of the amplifying loop is subtracted and replaced by the evolution of the input field after propagating in the ring (typically, half circumference).

In a more formal description, we define two space coordinates $X_{(a,b)}$ for the two cavities, with coupling points at $X_{(a,b)} = 0$ and $X_{(a,b)} = L_{(a,b)}^{(1)}$. The minimum propagation path of a pulse within the micro-cavity is $L_a = L_a^{(1)} + L_a^{(2)}$, which corresponds to a round-trip time $T_a$. Because there is no physical connection between points $X_b = L_b^{(1)}$ and $X_b = 0$ of the gain loop, a pulse propagating in the main-cavity necessarily travels the micro-cavity section $L_a^{(2)}$ so that the minimum propagation path is, in this case, $L_b^{(1)} + L_a^{(2)}$.

Therefore, in our model it is convenient to consider the main-cavity as closed with an adjusted round-trip time $T_b$, *o*btained by adding an auxiliary length $L_b^{(2)} \approx L_a^{(2)}$ between the points $X_b = L_b^{(1)}$ and $X_b = 0$ (Fig. S1a dashed black line) so that the total length of the cavity is, consequently, $L_b = L_b^{(1)} + L_b^{(2)}$. The length $L_b^{(2)}$ will be exactly defined below and we will properly configure the optical field at the main-cavity input point $X_b = 0$, so that it will depend only on the field in the micro-resonator at the point $X_a = L_a$.

We consider $T_b \gg T_a$ and define an integer $M$ which approximately represents the ratio between the two round-trips. Such a quantity allows us to count how many modes of the long cavity are contained in the free-spectral range (FSR) of the micro-cavity, Fig. S1b. Specifically, we have

$$F_a = (M - \delta)F_b, \tag{S1}$$

with the FSR given by $F_{(a,b)} = T_{(a,b)}^{-1}$ for the two cases. The variable $|\delta| < 1/2$ allows for modelling non-commensurate loops and represents the FSR detuning, as in Eq. (1) in the main text.

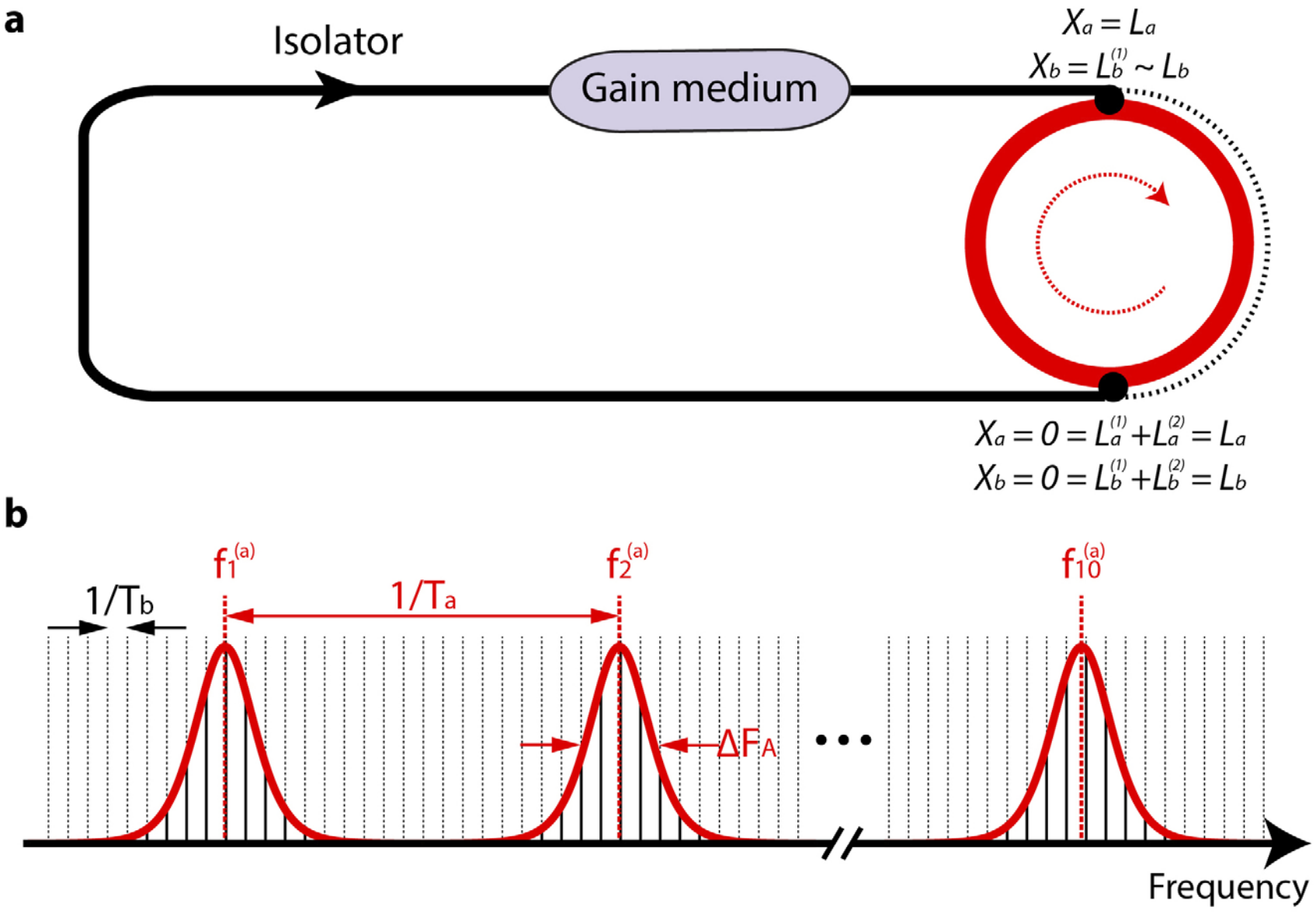

**Figure S1. Principle of operation. a.** Scheme of the nested, travelling-wave cavities configuration: a Kerr micro-cavity ('a' – red loop, with length $L_a = L_a^{(1)} + L_a^{(2)}$), is nested in an amplifying main cavity ('b' – black line, with length $L_b = L_b^{(1)} + L_b^{(2)}$). We define two space coordinates $X_{(a,b)}$ for the two cavities, with coupling points at $X_{(a,b)} = 0$ and $X_{(a,b)} = L_{(a,b)}^{(1)}$. **b.** Scheme of the modes interaction, with the resonances of the micro-cavity (red) and modes of the main-cavity (black). The Kerr micro-cavity resonances possess a linewidth $\Delta F_A$ and each mode (of index $m$) a central frequency $f_m^{(a)}$.

The optical fields in the two cavities $A(T, X_a)$ and $B(T, X_b)$ are expressed in [$\sqrt{\mathrm{W}}$] and are slowly varying in time $T$, defined in [s].

It is convenient to expand such functions in a set of cavity modes $a_m(T)$ and $b_n(T)$, defined in [$\sqrt{\mathrm{J}}$] as usually done in coupled-mode theory [1] as

$$A(T, X_a) = \sum_{m=-\infty}^{\infty} \frac{a_m(T)}{\sqrt{T_a}} \exp\left[2\pi i m \frac{X_a}{L_a}\right], \tag{S2}$$

$$B(T, X_b) = \sum_{n=-\infty}^{\infty} \frac{b_n(T)}{\sqrt{T_b}} \exp\left[2\pi i n \frac{X_b}{L_b}\right]. \tag{S3}$$

In this way, we can study their dynamics using a set of very general, standard coupled-mode equations [1] for all the interacting modes, which can be generalised for a large class of resonator types. Specifically, our full model, which accounts for the whole dynamic, reads as

$$\frac{da_m}{dT} = -\left[2\pi\, i\, f_m^{(a)} + \pi\, \Delta F_A + i\frac{v_a\beta_a}{2}(2\,\pi f_m^{(a)})^2\right] a_m - \frac{i\gamma v_a}{T_a}\sum_{j,p,l} \delta_{j-m,p-l}\, a_j a_l a_p^* + \frac{S_a}{\sqrt{T_a}}, \quad \text{(S4)}$$

$$\frac{db_n}{dT} = -\left[2\pi\, i\, f_n^{(b)} - G\, v_b + \left(i\frac{v_b\beta_b}{2} + \frac{1}{(2\pi\,\Delta F_F)^2 T_b}\right)(2\,\pi\, f_n^{(b)})^2\right] b_n + \frac{S_b}{\sqrt{T_b}}, \quad \text{S5)}$$

Here $\delta_{i,j}$ is the Kronecker delta. Every $a_m$or $b_n$ mode oscillates at the frequencies

$$f_m^{(a)} = m\, F_a, \text{ or} \quad \text{(S6)}$$

$$f_n^{(b)} = (n - \Delta)F_b, \quad \text{(S7)}$$

respectively. With this definition, we fix the central mode of the micro-cavity $a_0$ to a frequency $f_0^{(a)} = 0$, while the frequency $f_0^{(b)} = -F_b\,\Delta$ of the central mode of the laser cavity $b_0(t)$ is detuned via the parameter $\Delta$, which is the normalised frequency offset with respect to the central frequency $f_0^{(a)} = 0$ of the micro-cavity. Such offset is considered as a small value, in agreement with Eq. (1) in the main text.

We consider the waveguide dispersions $\beta_{(a,b)}$, in [$s^2m^{-1}$], the gain of the amplifying loop $G$, in [$m^{-1}$], the bandwidth $\Delta F_F$ of a pass band filter in [Hz] and the Kerr waveguide coefficient $\gamma$, in $W^{-1}m^{-1}$ [2]. The parameter $\Delta F_A$ in [Hz] is the -3 dB linewidth of the micro-cavity resonance [1].

We can now discuss the source terms $S_a$ and $S_b$, here expressed in [$\sqrt{W}$]. The cavities are coupled by means of the dimensionless coefficient $\sqrt{\theta}$ that we consider to be the same for the two ports, where $\theta$ is the ratio between the input and transmitted power in the coupler. Because the coupling losses are dominant in the micro-resonator, such a coefficient is related to the micro-cavity linewidth by $\theta = \pi\,\Delta F_A T_a$. [1].

The field coupled in the micro Kerr cavity, which is added to its intra-cavity field at the point $X_a = 0$, is the output field of the main-cavity in $X_b = L_b^{(1)}$, resulting in [1]

$$S_a = \sqrt{\theta}B\big(T, L_b^{(1)}\big) = \sqrt{\theta}\sum_{n=-\infty}^{\infty}\frac{b_n\,(T)}{\sqrt{T_b}}\exp\left[2\pi i n\frac{L_b - L_b^{(2)}}{L_b}\right] = \sqrt{\theta}\sum_{n=-\infty}^{\infty}\frac{b_n\,(T)}{\sqrt{T_b}}\exp\left[-2\pi i n\frac{L_a^{(1)}}{M L_a}\right], \quad \text{(S8)}$$

where we have set $L_b^{(2)}/L_b = L_a^{(1)}/(M\,L_a)$. Such an equation will simplify the calculation later.

The definition of the source term $S_b$ for the main-cavity needs to take into account that the main amplifying cavity is modelled as an effective open loop because in our experimental model there is no physical connection between the points $X_b = L_b^{(1)}$ and $X_b = L_b$ (black dashed line in Fig. S1a), as discussed at the beginning of this section. Such a modelling can be done imposing that the input field in the amplifying loop at $X_b = 0$ is only dependent on the micro-cavity coupled field, i.e. $B(t,0) = \sqrt{\theta}A(t, L_a^{(1)})$. From a practical point of view, this requires subtracting the field $B(t, L_b)$ propagating in the main-cavity just before the coupling point $X_b = 0$ in the source term,

$$S_b = \sqrt{\theta}A\big(T, L_a^{(1)}\big) - B(T, L_b) = \sum_{m=-\infty}^{\infty}\sqrt{\theta}\frac{a_m\,(T)}{\sqrt{T_a}}\exp\left[2\pi i m\frac{L_a^{(1)}}{L_a}\right] - \frac{b_m\,(T)}{\sqrt{T_b}}. \quad \text{(S9)}$$

Eqs. (S1)-(S9) define the full model which can be used to simulate the complete dynamic of the system and it is equivalent to the model described in Ref. [3]. Mean-field models, however, usually allow for better understanding of the physical dynamics and, for this reason, we will reduce Eqs. (S1)-(S9) with standard phase-matching considerations [2] to a set of partial differential equations.

In principle, Eqs. (S8) and (S9) couple all the $b_n$ modes, both together and to every $a_m$ mode. Practically, however, only modes with similar frequencies will interact [1]: the strongest interaction with the Kerr cavity frequency $f_m^{(a)}$ will occur for the main-cavity frequencies $f_n^{(b)}$ falling within the micro-cavity linewidth $\Delta F_A$:

$$\left|f_n^{(b)} - f_m^{(a)}\right| < \Delta F_A. \qquad \text{(S10)}$$

It is, therefore, important to express the main-cavity frequencies $f_n^{(b)}$ in terms of the Kerr cavity frequencies $f_m^{(a)}$ to allow a better comparison between the interacting frequencies. Considering that every FSR of the micro-cavity contains almost $M$ $f_n^{(b)}$ frequencies (see Eq. (S1) and Fig. S1b), we express the index $n$ of the $b_n$ set as

$$n = m\,M + q, \qquad \text{(S11)}$$

where $m$ spans all the integer spectrum while $|q| < M$. Using the definition of the integer $M$ given in Eq. (S1), we get

$$f_n^{(b)} = (m\,M + q - \Delta)F_b = m\,F_a + (q - \Delta + m\,\delta)F_b = f_m^{(a)} - (\Delta - q - m\,\delta)F_b. \qquad \text{(S12)}$$

The index $m$ can now be used to refer the main-cavity mode directly to the frequency $f_m^{(a)}$ of the micro-cavity. The parameter $q$ is, conversely, a relative index referring to the frequency of the $a_m$ mode. Following our discussion in the main text, it defines the order of the *super-mode*: in particular, $q=0$ is associated to the leading-order mode. In general, $q$ selects a set of equally-spaced modes of the main-cavity. When the FSR of the two cavities is commensurate, i.e. $\delta = 0$, such modes are exactly set apart by the micro-cavity FSR $F_a$. The super-modes and, practically, the index $q$ are very useful to select the interacting frequencies, although the number of modes necessary for correctly describing the interaction is difficult to determine *a priori*. We can, however, define an integer $N$ and keep only the super-modes with $|q| < N$. We expect that the number $2N + 1$ of main-cavity modes per micro-cavity resonance will be around the order of magnitude of the number of main-cavity modes per micro-cavity lines $\Delta F_A / F_b$ and, in any case, much smaller than the ratio $M$ between the two cavities' FSRs (Eq. (S1)). The validity of the solution can be simply checked *a posteriori* by testing the model for increasing $N$, as we will do in the following.

We can now look into a field expression for a single super-mode and, to this aim, we group together all the modes $b_n = b_{mM+q}$ with the same $q$ and Fourier transform them in the space $X_a$ of the micro-cavity

$$B_q(T, X_a) = \sum_{m=-\infty}^{\infty} \frac{b_{m\,M+q}(T)}{\sqrt{T_b}} \exp\left[2\pi i m\, \frac{\left(X_a - L_a^{(1)}\right)}{L_a}\right]. \qquad \text{(S12)}$$

Without loss of generality, we have centred the mode in $L_a^{(1)}$ because it is convenient in the following substitutions.

Now that both the micro- and main-cavity fields are defined in the same space, we derive a mean-field system via multiplying Eqs. (S4) and (S5) by the terms $\exp[2\pi i m\, X_a L_a^{-1}]$ and $\exp\left[2\pi i m\left(X_a - L_a^{(1)}\right)L_a^{-1}\right]$, respectively. We will then keep only a small number $2N + 1 << M$ of main-cavity modes per micro-cavity resonance: the mode $b_{m\,M+q}$ will interact with the mode $a_m$ only if $|q| < N$. Summing up for every $m$ and, by using Eqs. (S2) and (S12), we get

$$T_a \frac{\partial A}{\partial T} + L_a \frac{\partial A}{\partial X_a} - L_a \frac{i\beta_a v_a^2}{2} \frac{\partial^2 A}{\partial X_a^2} + i\gamma\, L_a |A|^2 A = -\theta A + \sqrt{\theta} \sum_{q=-N}^{N} B_q, \qquad \text{(S11)}$$

$$T_b \frac{\partial B_q}{\partial T} + M\, L_a \frac{\partial B_q}{\partial X_a} - L_b \frac{i\beta_b v_a^2}{2} \frac{\partial^2 B_q}{\partial X_a^2} - 2\pi i (\Delta - \mathrm{q}) B_q = \frac{v_a^2}{(2\pi\, \Delta F_F)^2} \frac{\partial^2 B_q}{\partial X_a^2} + G L_b\, B_q - \sum_{p=-N}^{N} B_p + \sqrt{\theta} A, \qquad \text{(S12)}$$

where we have neglected the contribution of the small detuning $q + \Delta + m\,\delta$ in the dispersion and of the small-phase terms $q/M$. The equations reported in the main text are found with the following normalisation. The propagating time is normalised against the main-cavity period in the frame moving with the pulses, $t = T\, T_b^{-1}$, while the fast cavity-time, $x = X_a L_a^{-1} - T\, T_b^{-1}$, is normalised against the micro-cavity roundtrip. We have $a = A\sqrt{\gamma v_a T_b}$, $b_q = B_q\, T_b \sqrt{\gamma v_a\, T_a^{-1}}$, $g = G\, L_b$, $\zeta_{(a,b)} = -\beta_{(a,b)} v_{(a,b)} T_b T_a^{-2}$, $\sigma = (2\pi\, \Delta F_F T_a)^{-2}$. Note that the coupling parameter $\kappa = \theta T_b F_a = \pi\, \Delta F_A\, T_b$ provides directly the number $\kappa\pi^{-1}$ of main-cavity modes per micro-ring resonance. Tables 1 and 2 in the Appendix summarise all the quantities used in the simulations.

Finally, we note that such a reduction could be also performed starting from a set of propagation equations, as modelled in Ref. [3], by applying a similar method to Ref. [4].

## Soliton Search and Analysis

We start from the general system used in the main text involving the coupled interaction of 2N+1 super-modes

$$\partial_t a + \frac{i\zeta_a}{2}\partial_{xx}a + i\,|a|^2 a = -\kappa a + \sqrt{\kappa}\sum_{q=-N}^{N} b_q, \tag{S13}$$

$$\partial_t b_q + \delta\partial_\tau b_q + \frac{i\zeta_b}{2}\partial_{xx}b_q - 2\pi\, i\,(\Delta - q)b_q = -\sigma\partial_{xx}b_q + g\, b_q - \sum_{p=-N}^{N} b_p + \sqrt{\kappa}a. \tag{S14}$$

The aim of this section is to find localised solutions of the kind $a(t,x) = a_S(x - vt)\exp[2\pi i\phi t]$, $b_q(t,x) = b_{q,S}(x - vt)\exp[2\pi i\phi t]$, where the variable $\phi$ is the normalised frequency of the pulse and $v$ the normalised velocity, as defined in the main text. We will do such a search via numerical integration. Before starting with the integration, it is convenient in any case to do some analytical considerations that will help us in defining the range of parameters where stable solitons exist. For this reason, we focus first on the case $N = 0$, involving only the mode $b_0(t,\tau)$: the study of the soliton behaviours in such a system is then associated directly with the leading cavity super-mode (experiencing the lowest coupling losses) responsible for driving the predominant dynamics and, thus, providing a general understanding of the full system. Eqs (S13) and (S14) simplify in a two-equations system

$$\partial_t a + \frac{i\zeta_a}{2}\partial_{xx}a + i\,|a|^2 a = -\kappa a + \sqrt{\kappa}b_0, \tag{S15}$$

$$\partial_t b_0 + \delta\partial_x b_0 + \frac{i\zeta_b}{2}\partial_{xx}b_0 - 2\,\pi i\Delta\, b_0 = -\sigma\partial_{xx}b_0 + (g-1)\, b_0 + \sqrt{\kappa}a, \tag{S16}$$

where we will assume $\zeta_a > 0$ to study bright localised states. [5] The system possesses strong similarities with the model for semiconductor solitons in frequency-selective feedback cavities and with dual-core fibre laser systems [6, 7], although it has the fundamental peculiarity of showing nonlinearity and gain in two different equations which, different from previous studies, allows the formation of broadband solutions in the Kerr cavity.

The existence of cavity-solitons generally requires bistability, i.e. the contemporary presence of two stationary states, a high-energy and a low-energy state (or background) upon which the soliton is formed. In the field of micro-combs, this has been largely discussed for Lugiato-Lefever temporal cavity solitons usually forming over a continuous-wave (CW) background; the Lugiato-Lefever equation, with the conventions used in this paper, reads as

$$\partial_t a + \frac{i\zeta_a}{2}\partial_{xx}a + i\,|a|^2 a - 2\,\pi i\Delta\, a = -\kappa a + \sqrt{\kappa}\,S, \tag{S17}$$

for the same dispersion $\zeta_a$, coupling constant $\kappa$ and field $a$ of the micro-cavity in (S15) and with $S$ as the external driving field, detuned from the coupled resonance by the frequency offset $\Delta$. The threshold for bistability of Eq. (S17) requires $\Delta < \kappa\sqrt{3}$ and $|S|^2 > 8/(3\sqrt{3})\,\kappa^2$, [8] which provides a minimum threshold excitation for soliton formation.

Conversely, in the system given by Eqs. (S15)-(S16), the two bistable stationary states are a CW state and a zero background. This is typical of homogeneous systems with gain, like in Refs. [6,7], where the null solution is admitted and usually exists in the full range of parameters, different from inhomogeneous, externally-driven systems, like in Eq. (S17). Eqs. (S15) and (S16) admit the CW solution $|a_S|^2 = 2\pi\left(\pm\sqrt{g\,(1-g)^{-1}}\,(1 - g + \kappa) - \Delta\right)$ with $\phi = -\Delta \pm (2\pi)^{-1}\sqrt{g(1-g)}$: its existence implies $0 < g < 1$. It is very important to stress that, different from Eq. (S17), the threshold for bistability is simply $g$ = 0: any excitation above the zero level can, in principle, provide a localised state.

The stability of the solitary states requires the stability of the low-energy stationary state, which here is the null solution. Such a state is always unstable for frequency offsets $|\Delta| < (2\pi)^{-1}\sqrt{g\,(1-g)^{-1}}|1-g+\kappa|$. Although this formula is, in principle, valid only for Eqs. (S15)-(S16), the instability regions of Eqs. (S13)-(S14) are well approximated by $|\Delta+q| < (2\pi)^{-1}\sqrt{g\,(1-g)^{-1}}|1-g+\kappa|$ for $N > 0$.

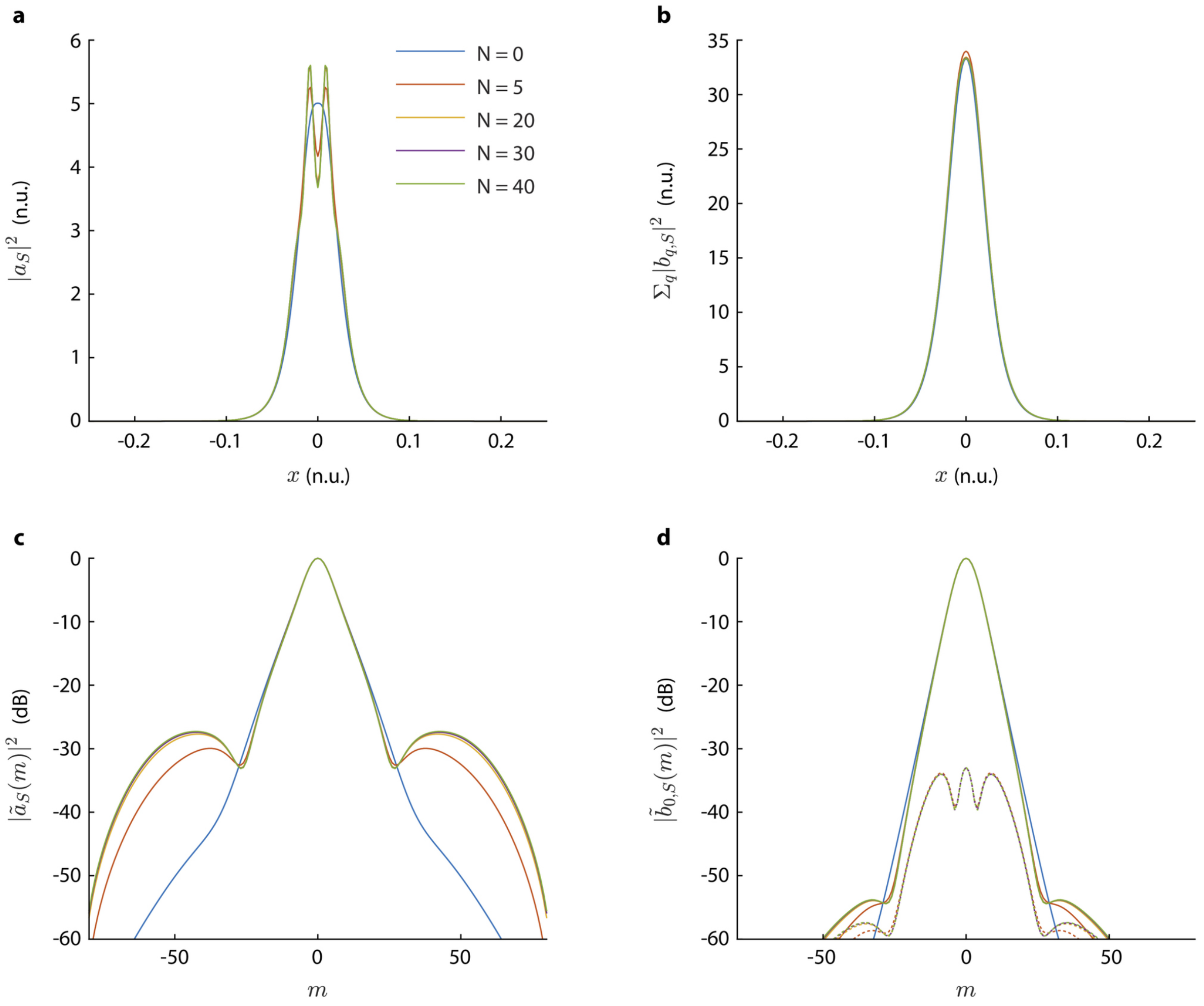

**Figure S2. Solitary profile for different approximations of the number of super-modes N.** Here the cases for *N = 0* (blue), *N = 5* (orange), *N = 20* (yellow), *N = 30* (purple) and *N = 40* (green) are reported. **a.** Time profiles of the field in the micro-cavity. **b.** Time profiles of the field in the main-cavity. **c.** Spectral intensity of the field in the micro-cavity. **d.** Spectral intensity of the field in the main-cavity. The spectral distribution of the first-order super-mode (*q = 1*) for the different cases is plotted with dashed lines in d. Here we have $\zeta_a = 1.25\times10^{-4}$, $\zeta_b = 3.5\times10^{-4}$, $\sigma = 1.5\times10^{-4}$, $\kappa = 2\pi$, $g = 0.14$, $\delta = 0$ and $\Delta = 0.4763$. It results in the following values: $\phi = -0.4735$ and $v = 0$.

These preliminary considerations help to restrict the range of parameters where interesting solitary solutions are expected. We now look numerically for localised states using the general system described by Eqs. (S13)-(S14). Starting from an appropriate guess solution, we use a numerical method (arc-length continuation [9]) for finding the soliton families of Eqs. (S13) and (S14). As discussed in the main text and reported in the Appendix, we use the parameters $\zeta_a = 1.25\times10^{-4}$, $\zeta_b = 3.5\times10^{-4}$, $\sigma = 1.5\times10^{-4}$ and $\kappa = 2\pi$, consistent with our experimental system.

Figure S2 reports a comparison of the solitary profiles obtained with $N$ = 0, $N$ = 5, $N$ = 20, $N$ = 30 and N = 40 for $g = 0.14$, $\delta = 0$ and $\Delta = 0.4763$. This first test aims to set the number of modes $N$ necessary for the numerical analysis. The difference from the case $N$ = 0 is mostly visible in the spectral intensity plot[1] of the field, showing that the dominant dynamic is related to the leading modes. The cases $N$ = 20, $N$ = 30 and $N$ = 40 are indiscernible over the entire spectral range. The case $N$ = 5 provides a good approximation in this range of parameters. Higher-order modes in the micro-resonator (Fig. S2d), which are all locked to the soliton frequency $\phi = -0.4735$, here only appear -30 dB below the leading order: this analysis shows, however, that they contribute to provide a better spectral description of the comb.

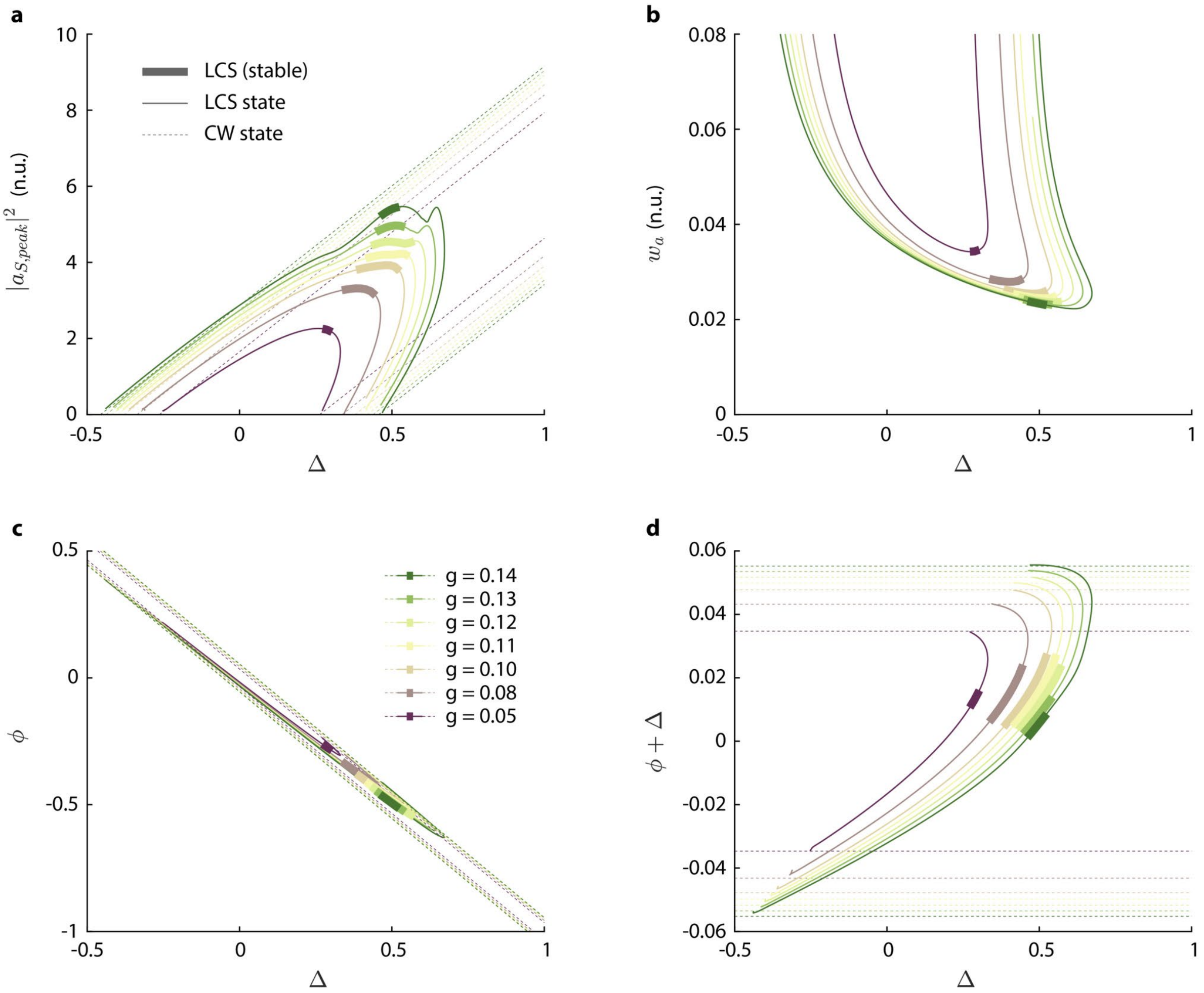

**Figure S3. Map of Soliton States.** Stationary solutions for $\zeta_a = 1.25 \times 10^{-4}$, $\zeta_b = 3.5 \times 10^{-4}$, $\sigma = 1.5 \times 10^{-4}$, $\kappa = 2\pi$, $\delta = 0$ for $g = 0.05, 0.08, 0.10, 0.11, 0.12, 0.13, 0.14$ (plot colour scale ranging from purple to green, respectively). Thick lines mark stable solutions. **a.** Peak intensity of the solitary field in the micro-cavity vs frequency cavity offset Δ (full lines) with corresponding two CW solutions (dashed lines) for $\delta = 0$. **b.** Pulse width of the micro-cavity intensity profile, defined as its standard deviation. **c.** Frequency of the soliton and of the CW solutions (solid and dashed lines respectively). **d.** Residual frequency $\phi + \Delta$.

A general understanding of the behaviour of such solitary solutions can be obtained by analysing the fundamental soliton family as a function of the normalised frequency offset Δ, FSR detuning $\delta$ and gain $g$. First, we find the families by continuation for $N$ = 5 and use standard linear perturbation analysis for defining their range of stability. Then, the stable solutions are propagated in a system with $N$ = 30 to test their stability with a more accurate approximation.

[1] Fourier transform of the solution is defined as $\tilde{a}_s(m) = \int a_s(x) e^{-2i\pi m x} dx$ where $m$ is the mode number.

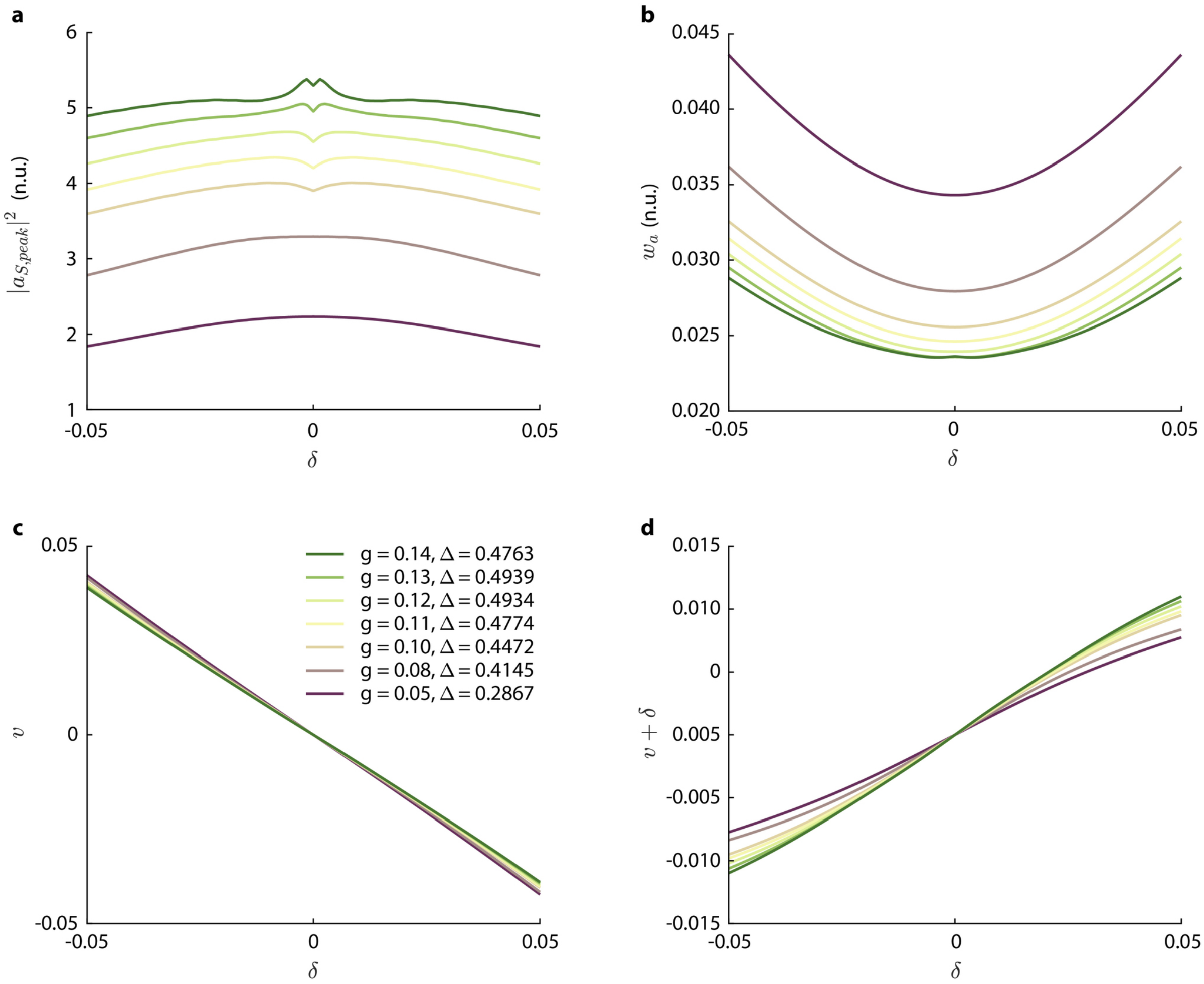

**Figure S4. Map of Walking Soliton States.** Walking stationary solutions for $\zeta_a = 1.25 \times 10^{-4}$, $\zeta_b = 3.5 \times 10^{-4}$, $\sigma = 1.5 \times 10^{-4}$, $\kappa = 2\pi$, for $g = 0.05, 0.08, 0.10, 0.11, 0.12, 0.13, 0.14$ (plot colour scale ranging from purple to green). Here Δ= 0.2867, 0.4145, 0.4472, 0.4774, 0.4934, 0.4939, 0.4763. **a.** Peak intensity of the solitary field in the micro-cavity vs velocity mismatch $\delta$. **b.** Pulse width. **c.** Velocity. **d.** Residual velocity $v + \delta$.

The results obtained for $\delta = 0$ by varying Δ are reported in Fig. S3a for $g$ ranging from 0.5 to 1.4. The peak intensity of $a_S(\tau)$ (i.e. for the micro-cavity field) is reported in Fig. S3a (full lines from purple to green), along with the peak intensity of the CW solutions, dashed lines. Stable solutions are marked with a thick line. Figure S3b depicts the solution pulse width $w_a$ defined as the standard deviation of the intensity. Stable solutions correspond to the narrowest pulses. The soliton frequency $\phi$ (solid lines) along with the frequency $\phi$ for the CW stationary states (dashed lines) are reported in Fig. S3c. The frequency is better visualised by $\phi + \Delta$ in Fig. S3d, which shows that the oscillating frequency $\phi$ of the localised states is distinct from the CW case.

After the analysis for stationary solitons, we look for so-called *walking* solitary solutions [8] (with a velocity $v$ that is not zero) versus $\delta$ for the same range of gain used in the previous case. The offset Δ is chosen at the lowest border of the stability region. The peak intensity, pulse width and velocity of the micro-cavity field $a_S(x)$ are reported in Fig. S4. Remarkably, the state remains stable in the explored range.

## Experimental Setup and Further Measurement Analysis

Figure S5 reports the experimental setup. In the results presented in Figs. 2 and 3 in the main text, the comb frequency lines have been finely characterised, with accuracies below the MHz scale, by intra-cavity laser-scanning spectroscopy [9] (Fig. S5), which also revealed the position of the comb lines with respect to the micro-cavity resonances. In this way, we could experimentally extract the variables $\phi$ and $\nu$ of the solitons, allowing a proper comparison with the theory and, thus, also obtain information on the presence of additional super-modes.

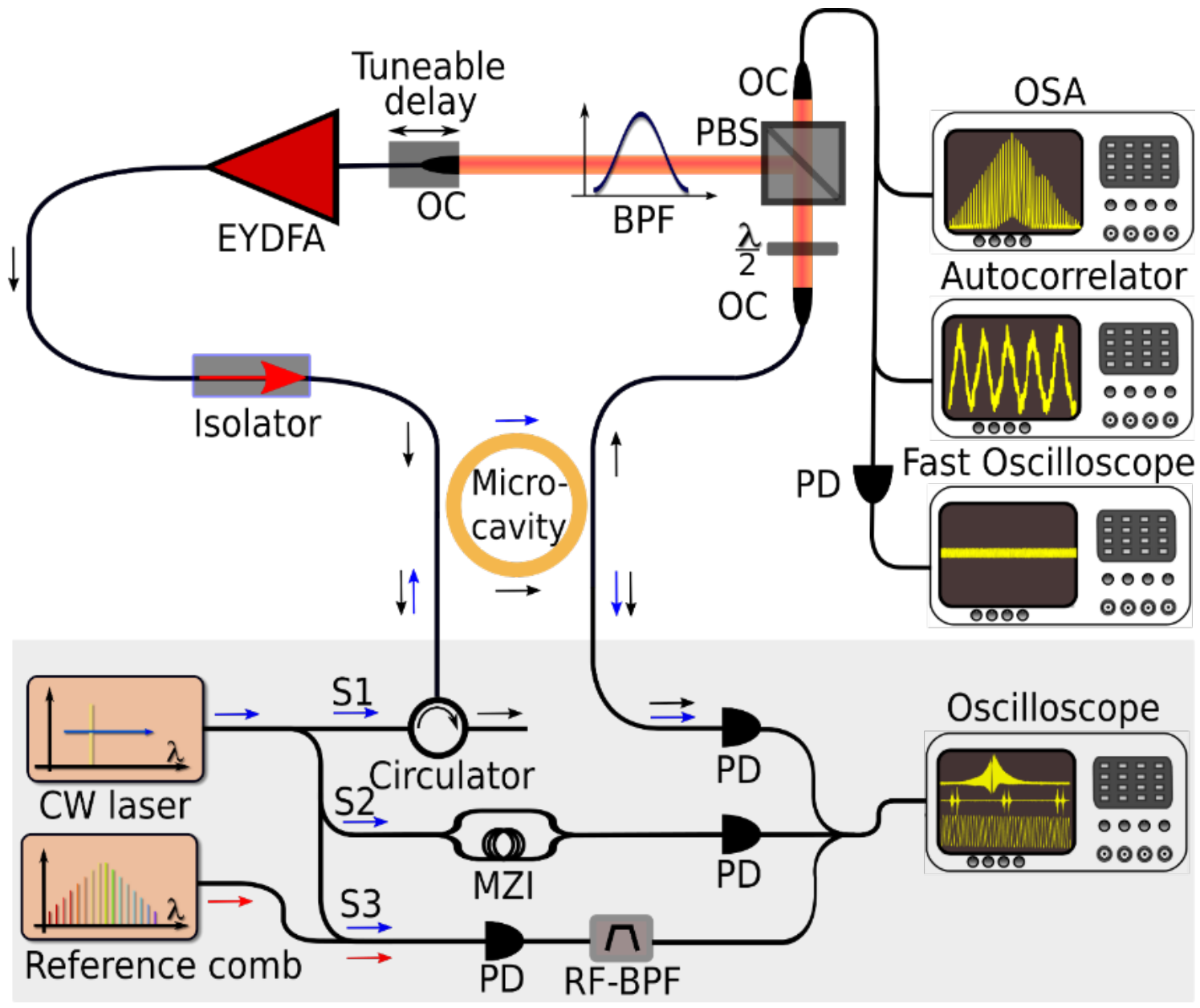

**Figure S5. Experimental set-up for micro-comb cavity-solitons' generation and laser-scanning spectroscopy detection of a hot resonator.** The set-up is composed of a nonlinear micro-resonator, an erbium-ytterbium co-doped fibre amplifier (EYDFA), an optical isolator, an optical bandpass filter (BPF), a tuneable delay line, a half-wave plate (λ/2), a polarising beam splitter (PBS) and two optical collimators (OCs). The output signals from the rejection port of the polarising beam splitter were detected with an optical spectrum analyser (OSA), an autocorrelator and a fast photodetector (PD) connected to an oscilloscope. For the laser spectroscopy, a scanning CW laser was split into three signals, S1, S2 and S3. S1 was used to probe the resonances' profile and oscillating micro-comb lines in the hot micro-resonator, while S2 and S3 were simultaneously used to perform frequency calibration of the scanning laser. This was achieved by propagating the external CW source in a Mach–Zehnder interferometer (MZI) and beating it with a reference comb (Menlo System). The resulting signal was passed through a radio frequency bandpass filter (RF-BPF) before detection.

The presence of such modes was also verified by extracting the radio frequency noise of the comb intensity, which is reported in the inset of Fig. S6a, b, along with the results also shown in Fig. 2 in the main text, reported here for convenience. The experiments correspond to an overtone generation (two solitons per round-trip) with repetition-rate twice the FSR of the micro-cavity.

Intra-cavity laser-scanning spectroscopy (Fig. S6b and d) also reveals a small, blue-detuned frequency found in the central resonances, yet absent in the comb wings. We attribute this perturbation to a small CW oscillation of the first order super-mode that produces a weak (-20 dB) beat-note at 77 MHz in the radio frequency spectrum (see Fig. S6a and c). In Fig. S6c, the radio frequency spectrum also shows a -20dB component at 6 MHz, which we attribute to the beating of the laser cavity-soliton with a small CW perturbation of the leading mode, resulting in well-distinguished frequencies.

The presence of additional components in higher-order modes that are not locked to the soliton solution and can coexist with it is well supported by the theoretical analysis. We tested the propagation of the calculated solution which matches the experiments, for Δ = 0.49; g = 0.1 and Δ = 0.47; g = 0.14, corresponding to the cases of Fig. S6a and b, respectively. The propagations are reported in Fig. S7a and d, (depicted by false colours in log scale), to better visualise the low energy component of the solution.

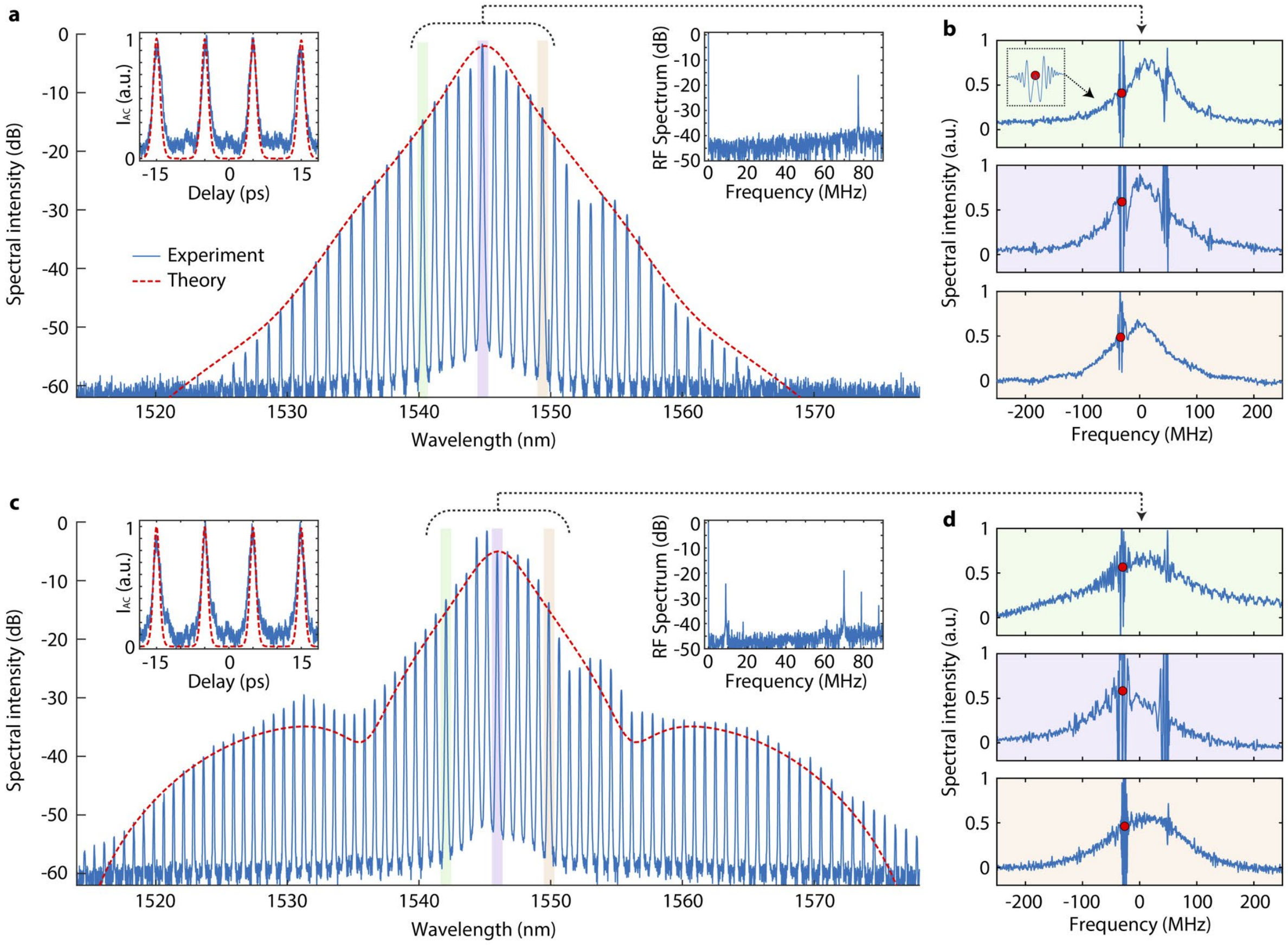

**Figure S6. (as in Fig. 2 of the main text). a.** Soliton generation with 20 mW intra-cavity power: spectrum (in logarithmic scale), autocorrelation (right inset) and radio frequency noise (left inset). The experimental measurements are plotted in blue and directly compared to the theoretical solitary state (red, fit parameters: Δ = 0.49; g= 0.1). **b.** Intra-cavity spectrum (blue), evidencing the lasing modes (red dots) within each micro-cavity resonance. The three plots correspond to different comb wavelengths, illustrated by the colour shading in panel a. **c,d.** The same measurements for a higher fibre gain, leading to a 30 mW intra-cavity power and a larger comb spectrum. Fit parameters are Δ =0.47; g= 0.14.

We propagate the two equidistant solitons, as observed in the experiments. Figure S6b and d, report an evaluation of the effective propagating frequency (normalised, as the detuning Δ, against the main-cavity FSR) at the peak ($x = 0.25$, blue line) and at the minimum of the propagating intensity ($x = 0$, orange line). Such a spectrum is calculated as the Fourier transform[2] along the propagating slow axis $t$ at those specific values of $x$. The frequency axis $\xi$ is centred along the soliton frequency $\phi$. This approach reveals the radio frequency spectrum of the solution and it is also used for the examples of Fig. 1 in the main text, allowing visualising the presence of unlocked higher-order super-modes.

Here our analysis shows that, in both cases, injecting the stationary states into the system also excites a very small background with the presence of unlocked components. Although such a background fades away in the propagation of the stationary states because they are stable in this range of parameters, this analysis shows that they can coexist with solitary propagation.

[2] Formally, $\tilde{a}(\xi, x) = \int a(t, x) e^{-2i\pi\xi t} dt$ where $\xi$ is the propagation (slow) axis frequency.

Very interestingly, given that such unlocked components of the higher-order modes correspond to the frequencies of the CW stationary states of the zero ($\xi = -0.05$) and first-order ($\xi = -1$) mode, we could find a direct match with the experimentally-observed radio frequency spectrum.

Note that the presence of perturbations with a frequency distinct from the stationary solution is in stark contrast to externally-driven micro-cavity systems, where all the co-existing solutions of the Lugiato-Lefever equation [8] (cavity-solitons and their perturbations) are intrinsically locked to the optical frequency of the driving field.

Finally, the input average energy to the micro-resonator, calculated as $\sum_{q=-N}^{N} \int |b_q(x)|^2 dx$ from the numerical fitting in Fig. S6a and b (or Fig. S7a and d), is 3.1 and 3.8 for the two cases, respectively. Compared to the formation threshold of Lugiato-Lefever bright solitons $8/(3\sqrt{3})\,\kappa^2 = 60.8$, they represent a fractional power of 5.1% and 6.3%, respectively.

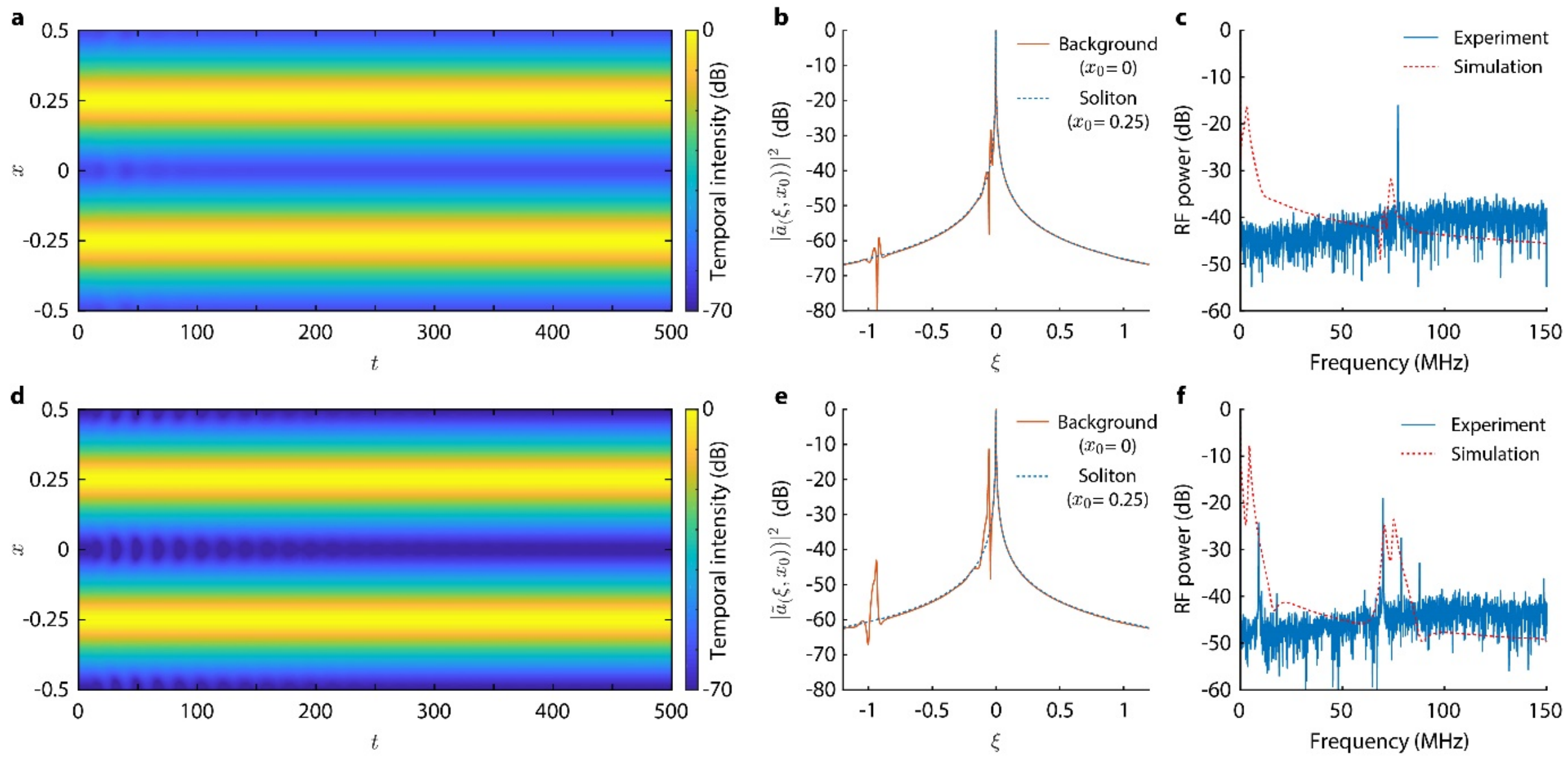

**Figure S7. Experimental propagation of soliton pulses.** Here $\zeta_a = 1.25 \times 10^{-4}$, $\zeta_b = 3.5 \times 10^{-4}$, $\sigma = 1.5 \times 10^{-4}$, $\kappa = 2\pi$, Δ=0.4, for $g = 0.1$ and 0.14 for a, b, c and d, e, f respectively. **a.** Propagation of two soliton pulses, intensity is displayed in false colours and logarithmic scale to better visualise the low energy components of the spectrum. **b.** Slow temporal scale (radio frequency) spectra obtained as the Fourier transform $\tilde{a}(\xi, x_0) = \int a(t, x_0) e^{-2i\pi\xi t} dt$ of the propagating solution at peak (blue, $x_0 = 0.25$) and tail (orange, $x_0 = 0$) of the pulse along the propagating time axis. The frequency axis is normalised with respect to the main-cavity FSR. **c.** Comparison of the experimental RF spectrum (blue) with the background RF spectrum calculated in the simulations as $\int |a(t, x_0)|^2 e^{-2i\pi\xi t} dt$ .

## Appendix

Table 1: Quantities Used in the Full Model

| Name | Symbol | Units or Values |
|---|---|---|
| **Coordinates** | $T, X$ | [s], [m] |
| **Fields** | $A, B$ | $[\sqrt{\mathrm{W}}]$ |
| **Periods, FSRs, Length and velocities** | $T_a = F_a^{-1} = L_a v_a^{-1}$, $T_b = F_b^{-1} = v_b L_b^{-1}$ $F_a = (M - \delta) F_b$. | $F_a = 48.9\mathrm{GHz}, F_b = 77\mathrm{MHz}$, $M \approx \frac{F_a}{F_b} = 635$ |
| **Dispersions** | $\beta_a, \beta_b$ | $\beta_a = -20\mathrm{ps}^2\mathrm{km}^{-1}$, $\beta_b = -60\mathrm{ps}^2\mathrm{km}^{-1}$ |
| **Coupling constant** | $\theta = \pi \Delta F_A T_a$ | $\Delta F_A = 150\mathrm{MHz}$, linewidth |
| **Gain Bandwidth** | $\Delta F_F$ | $\Delta F_F = 650$ MHz |
| **Gain** | $G$ | $[\mathrm{m}^{-1}]$ |
| **Kerr Waveguide Coefficient** | $\gamma$ | $[\mathrm{m}^{-1}\mathrm{W}^{-2}]$ |

Table 2: Conversions between the Full Model and Normalised Model

| Name | Symbol | Value |
|---|---|---|
| **Time Coordinate** | $t = \frac{T}{T_b}$ | - |
| **Space Coordinate** | $x = \frac{X}{L_a} - \frac{T}{T_b}$ | - |
| **Micro-Cavity Field** | $a\,(t, x) = A(T, X)\sqrt{\gamma T_b v_a}$ | - |
| **Gain-Cavity Field** | $b_q\,(t, x) = B_q(T, X) T_b \sqrt{\frac{\gamma v_a}{T_a}}$ | - |
| **Gain** | $g = G L_b$ | $0 < g < 1$ |
| **Dispersion ring** | $\zeta_a = -\frac{T_b v_a \beta_a}{T_a^2}$ | 0.000124 |
| **Dispersion fibre** | $\zeta_b = -\frac{T_b v_b \beta_b}{T_a^2}$ | 0.000372 |
| **Gain dispersion** | $\sigma = (2\pi T_a \Delta F_F)^{-2}$ | 0.000143 |
| **Coupling Constant** | $\kappa = \theta F_a T_b = \pi \Delta F_A T_b$ | $2\pi \approx 6.12$ |